\documentclass[aps,pre,preprint,groupedaddress]{revtex4}

\def\av#1{\left\langle#1\right\rangle}

\usepackage{graphicx}
\usepackage{bm}
\usepackage{graphicx}
\usepackage{subfigure}
\usepackage{latexsym}
\usepackage{amstext,amssymb,amsfonts}
\usepackage{epsfig}
\usepackage{color}

\begin{document}

\title{Efficiency of message transmission using biased random walks in complex networks in the presence of traps}

\author{Loukas Skarpalezos}

\affiliation{Department of Physics, University of Thessaloniki, 54124 Thessaloniki, Greece}

\author{Aristotelis Kittas}

\affiliation{Department of Physics, University of Thessaloniki, 54124 Thessaloniki, Greece}

\author{Panos Argyrakis}

\affiliation{Department of Physics, University of Thessaloniki, 54124 Thessaloniki, Greece}

\author{Reuven Cohen}

\affiliation{Department of Mathematics, Bar-Ilan University, Ramat-Gan 5290002, Israel}

\author{Shlomo Havlin}

\affiliation{Department of Physics, Bar-Ilan University, Ramat-Gan 52900, Israel}

\date{\today}

\begin{abstract}
We study the problem of a particle/message that travels as a biased random walk towards a target node in a network in the presence of traps. The bias is represented as the probability $p$ of the particle to travel along the shortest path to the target node. The efficiency of the transmission process is expressed through the fraction $f_g$ of particles that succeed to reach the target without being trapped. By relating $f_g$ with the number $S$ of nodes visited before reaching the target, we firstly show that, for the unbiased random walk, $f_g$ is inversely proportional to both the concentration $c$ of traps and the size $N$ of the network. For the case of biased walks, a simple approximation of $S$ provides an analytical solution that describes well the behavior of $f_g$, especially for $p>0.5$. Also, it is shown that for a given value of the bias $p$, when the concentration of traps is less than a threshold value equal to the inverse of the Mean First Passage Time (MFPT) between two randomly chosen nodes of the network, the efficiency of transmission is unaffected by the presence of traps and almost all the particles arrive at the target. As a consequence, for a given concentration of traps, we can estimate the minimum bias that is needed to have unaffected transmission, especially in the case of Random Regular (RR), Erd\H{o}s-R\'{e}nyi (ER) and Scale-Free (SF) networks, where an exact expression (RR and ER) or an upper bound (SF) of the MFPT is known analytically. We also study analytically and numerically, the fraction $f_g$ of particles that reach the target on SF networks, where a single trap is placed on the highest degree node. For the unbiased random walk, we find that $f_g \sim N^{-1/(\gamma-1)}$, where $\gamma$ is the power law exponent of the SF network.
\end{abstract}


\maketitle


\section{Introduction}

An important process usually associated with random walks is trapping. Trapping reactions have been widely studied as part of the general reaction-diffusion scheme. The trapping reaction can be formulated as: $A+T\rightarrow T$, where $T$ is a static trap and $A$ is a diffusing species that is annihilated irreversibly  when it comes in contact with the trap. The problem has been studied in a variety of geometries, such as regular lattices, in fractal spaces \cite{weiss_random, hollander_problems,bunde_prl78,havlin_physa169,donsker_com32} and recently, in small-world \cite{jasch_pre64}, Erd\H{o}s-R\'{e}nyi \cite{kittas_epl84,kittas_pre80}, and scale-free networks \cite{gallos_pre70,kittas_epl84,kittas_pre80}.

Such trapping processes can be related to the efficiency of message transmission in networks in the presence of one or more traps. The problem can be regarded of as an analogue for the propagation of information in certain communication networks in the form of packets. This follows since in some cases data packets traverse the network in a random fashion (for example, in wireless sensor networks \cite{avin_query}, ad-hoc networks \cite{yossef_mobi} and peer-to-peer networks \cite{gkantsidis_peer}). A trap acts as a node which is malfunctioning and where information is lost, e.g. like a router which can receive but not transmit data due to a malfunction or an e-mail server unable to forward incoming mails. However, information about the structure of the network might provide an opportunity to send a message with a bias towards the target. In \cite{skarpa_pre}, it was shown that using such a bias significantly reduces the time of diffusion from the source to the target. In ER networks there exists a threshold value of the bias parameter delimiting a power law and a logarithmic scaling of the MFPT with the size $N$ of the network. In SF networks, the scaling of the MFPT with network dimensions is always less than a power of $\log N$, i.e the gain of time is very important in SF networks, even for a small value of the bias parameter. When a certain number of nodes or a single important node, such as a hub, lose their functionality and act like traps, since some messages going threw these nodes will never arrive to their target, it is of interest to know the percentage of messages that are lost. It is, therefore, important to see if and how applying a bias towards the target nodes can help in saving messages.

\section{Model and Methods}

Our model can be described as follows: Messages start to be transmitted from a source node, with a target node as a destination. Both source and target nodes are chosen randomly from the total number of network nodes, but always on the giant connected component. Messages may diffuse on the network randomly, or with a bias toward the target node. Traps act as malfunctioning nodes in which information is lost (e.g., a router which cannot transmit data due to some failure).  We use Monte Carlo computer simulations. As described in \cite{skarpa_pre}, at each step, the particle travels either on a shortest path towards the target, with a probability $0 \leq p \leq 1$, or randomly to one of its node neighbors with a probability $1-p$. The particle may be trapped before it arrives on the target. The efficiency of transmission from the source to the target node is given by $f_g$, which is the fraction of particles that arrive from the source to the target. Obviously, the closer this value is to 1, the more efficient the transmission process is. 

The algorithm of the model may be described as follows:
\begin{enumerate}
	\item A pair of random nodes is selected as source and target nodes. A particle begins moving on the network from the source node.
	\item At each time step the particle hops to an adjacent node along the shortest path (between source and target) with probability $p$, or to a random node (including these on the shortest path) with probability $1-p$
	\item If the particle is trapped or arrives on target go to step 1, else go to step 2 (i.e. perform another time step)
\end{enumerate}
We consider the process only on the largest cluster of the network. We perform a total of $10^5$ realizations using 100 different networks, while performing walks between 1000 pairs of random source-target nodes on each network.

\section{Results for concentration of traps}

A key quantity for solving the problem of trapping is the number of distinct nodes $S$ visited before reaching the target. Indeed, $f_g$ is in reality the survival probability of the particles until meeting the target. If we denote by $q_{S=m}$ the probability that the number of distinct nodes visited before reaching the target is equal to $m$, then the fraction $f_g$ of particles that succeed to reach the target without being trapped can be expressed exactly as the following sum:
\begin{equation}
\label{eq:fgexact}
 f_g=\sum_{m=1}^{m=N}{q_{S=m}(1-c)^{m-1}}.
\end{equation}

For the unbiased random walk, since each node of the network (except the source) has the same probability of being the target, $f_g$ can be easily estimated:

\begin{equation}
\label{eq:fgrw}
 f_g=\sum_{m=1}^{m=N}{q_{S=m}(1-c)^{m-1}} \approx \frac{1}{N-1}\int_{m=1}^{m=N}{\mathrm{e}^{-c(m-1)}dm}\approx \frac{1}{Nc}.
\end{equation}

\begin{figure}[htbp]
    \begin{center}
    \subfigure[]{
    \includegraphics[width=7cm]{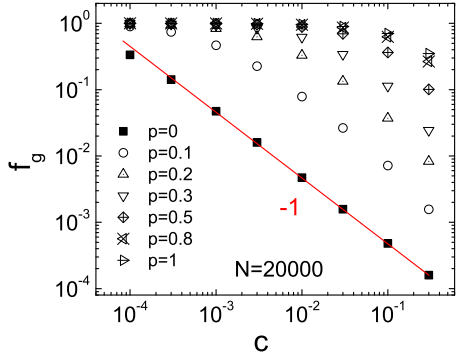}
    \label{ERfg1:a}}
    \subfigure[]{
    \includegraphics[width=7cm]{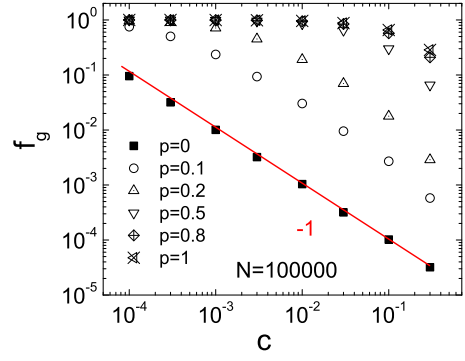}
    \label{ERfg1:b}}
		\subfigure[]{
    \includegraphics[width=7cm]{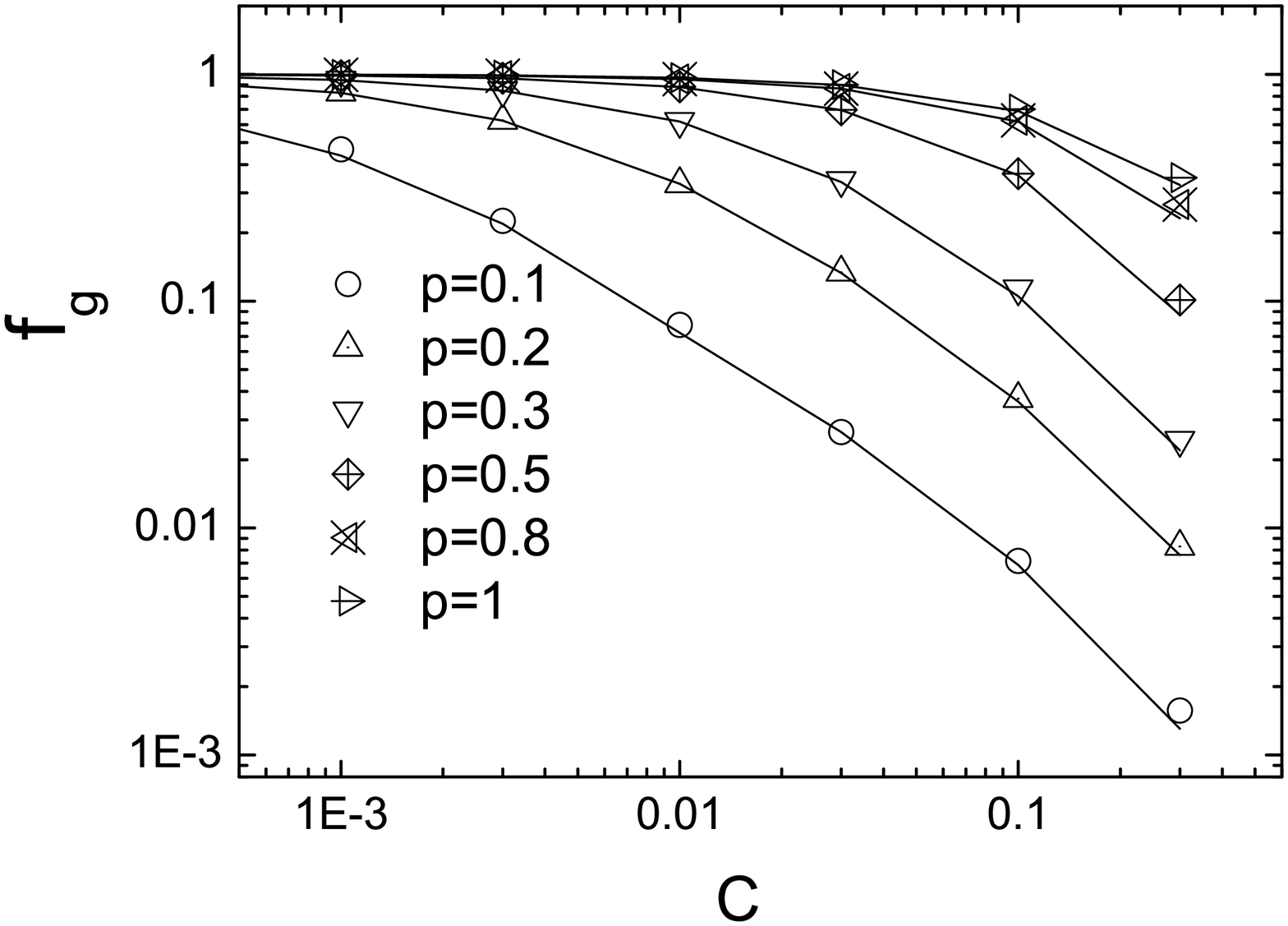}
    \label{ERfg1:c}}
				\subfigure[]{
    \includegraphics[width=7cm]{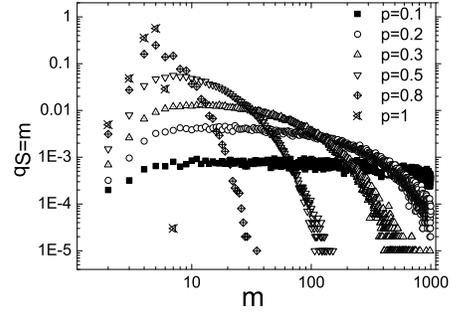}
    \label{ERfg1:d}}
    \end{center}
		\caption{(Color online) Log-log plot of $f_g$ vs $c$ for various values of $p$ in ER networks with $\av{k}=10$ for (a) $N=20000$ and (b) $N=100000$ respectively. (c) Comparison of  results for $N=20000$ (full symbols) with results from Eq. (\ref{eq:fgexact}) (continuous lines) where (d) the probability $q_{S=m}$ for having $m$  distinct nodes visited for different $p$ values is obtained from simulations without traps on similar networks.}
\label{ERfg1}		
\end{figure}

\begin{figure}
\begin{center}
\includegraphics[width=12cm]{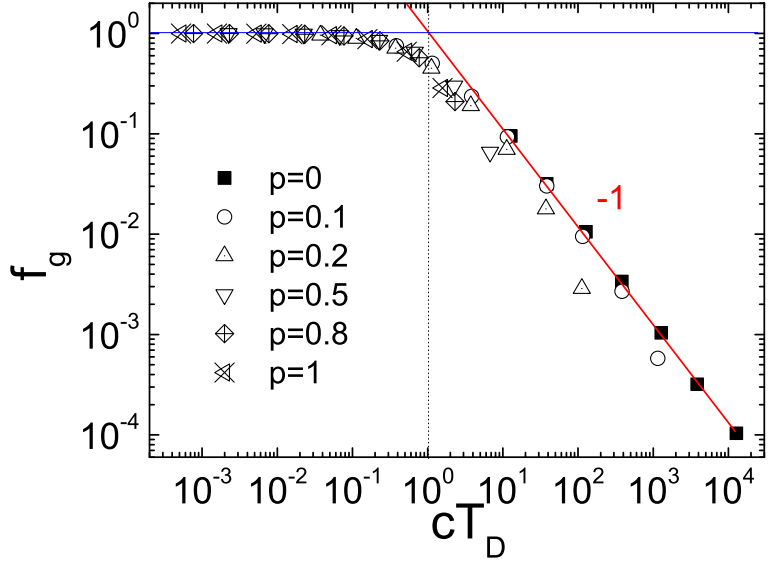}
\end{center}
\caption{(Color online) Log-log plot of $f_g$ vs $c T_D$ for ER networks with $\av{k}=10$ where $T_D$ is the MFPT in the case with no traps calculated from Eqs. (3) and (4). Here, we vary the trap concentration $c$ and $N=100000$, for various values of $p$.}
\label{ERfg2}
\end{figure}

Firstly, we plot the transmission efficiency of the biased random walk in the presence of a concentration $c$ of traps in ER networks obtained by simulations. In Figs. \ref{ERfg1:a}, \ref{ERfg1:b} we plot the fraction of particles that arrive on targets $f_g$, as a function of the concentration of traps $c$ for different values of the bias parameters in ER networks. Eq. (\ref{eq:fgexact}) is tested in Fig. \ref{ERfg1:c} by plotting both the results from simulations with traps (full symbols) and the evaluation of Eq. (\ref{eq:fgexact}) (continuous lines) where $q_{S=m}$ is obtained from simulations without traps, see Fig. \ref{ERfg1:d}. As expected from Eq. (\ref{eq:fgrw}), we see in Fig. \ref{ERfg1} that the probability that the unbiased particle will arrive at the target before meeting a trap is inversely proportional to both the size of the system and the concentration of traps. However, the situation is different for biased random walks. Indeed, in Fig. \ref{ERfg1} we see that, for low values of $c$, the particle may not hit the traps even with a relatively very small bias, and practically all particles arrive their target. For higher concentration of traps one needs stronger bias to survive and reach the target. This follows, because a stronger bias towards the target makes the path length shorter which helps to encounter less possible traps. Note that for small bias $p$, and for intermediate range of $c$, $f_g$ is almost parallel to $f_g$ at $p=0$. This suggests that in this range also the survival, $f_g$, of biased particles is inverse proportional to $c$. As seen in  Fig. \ref{ERfg1}, the concentration $c$ above which $f_g$ starts to behave as $1/c$ decreases  with decreasing bias. We hypothesize that this crossover is due to two competing time  scales. One is the typical time a walker does not encounter  traps, $1/c$, and the second is the mean first passage time (MFPT) from source to target without the presence of traps. We test this hypothesis in the scaling of Fig. \ref{ERfg2}.

In Fig.\ref{ERfg2}, we plot the same data as in Fig. \ref{ERfg1} but instead of $c$ on the $x$ axes we use $c T_D$, where $T_D$ is the MFPT from the source to the target in the case with no traps calculated from Eqs. (\ref{eq:solbis}) and (\ref{eq:distance}). Indeed, except from very high concentration of traps (order 1) the curves collapse into a single curve. This collapse clearly suggests the above discussed two distinct time scales and a crossover point corresponding to a threshold concentration $c_{th} = T_D^{-1}$. 

For $c < c_{th}$ the concentration of traps does not affect at all the transmission efficiency and practically $f_g\approx 1$. This seems reasonable since $1/c$ can be considered as the mean time spent between traps and if the time to reach the target is less than the time to diffuse from one trap to the other then the particle practically will not meet the traps. This result gives us a relation that connects directly the concentration of traps in a network of size $N$ with the value of the minimum bias parameter $p_m$ that is needed to have completely full efficiency. In \cite{skarpa_pre}, the MFPT was found analytically for RR and ER networks for every value of the bias parameter, see Eqs. (\ref{eq:solbis}) and (\ref{eq:distance}). Thus, in both RR and ER networks with a concentration c of traps, $p_m$ can be found easily by solving the following equation: 

\begin{equation}
\label{eq:solbis}
T_D =\frac{D}{2p'-1}+\frac{1-p'}{(2p'-1)^2}\left[\left(\frac{1-p'}{p'}\right)^D-1\right]\ =\frac{1}{c} .
\end{equation}
where $p'=p+(1-p)/\av{k}$ and
\begin{equation}
\label{eq:distance}
D=\frac{\ln\left(1+(\av{k}-2)(N-1)/\av{k}\right)}{\ln (\av{k}-1)}\approx\frac{\ln\left((\av{k}-2)N/\av{k}\right)}{\ln (\av{k}-1)}
\end{equation}

Furthermore, in Figs. \ref{ERfg1} and \ref{ERfg2} we see that for $c > c_{th}(p)$ the transmission efficiency is strongly affected by traps and all the curves follow the unbiased one that has a slope of $-1$ in the log-log plot except for very large c of order 1.

\begin{figure}[htbp]
    \begin{center}
    \subfigure[]{
    \includegraphics[width=7cm]{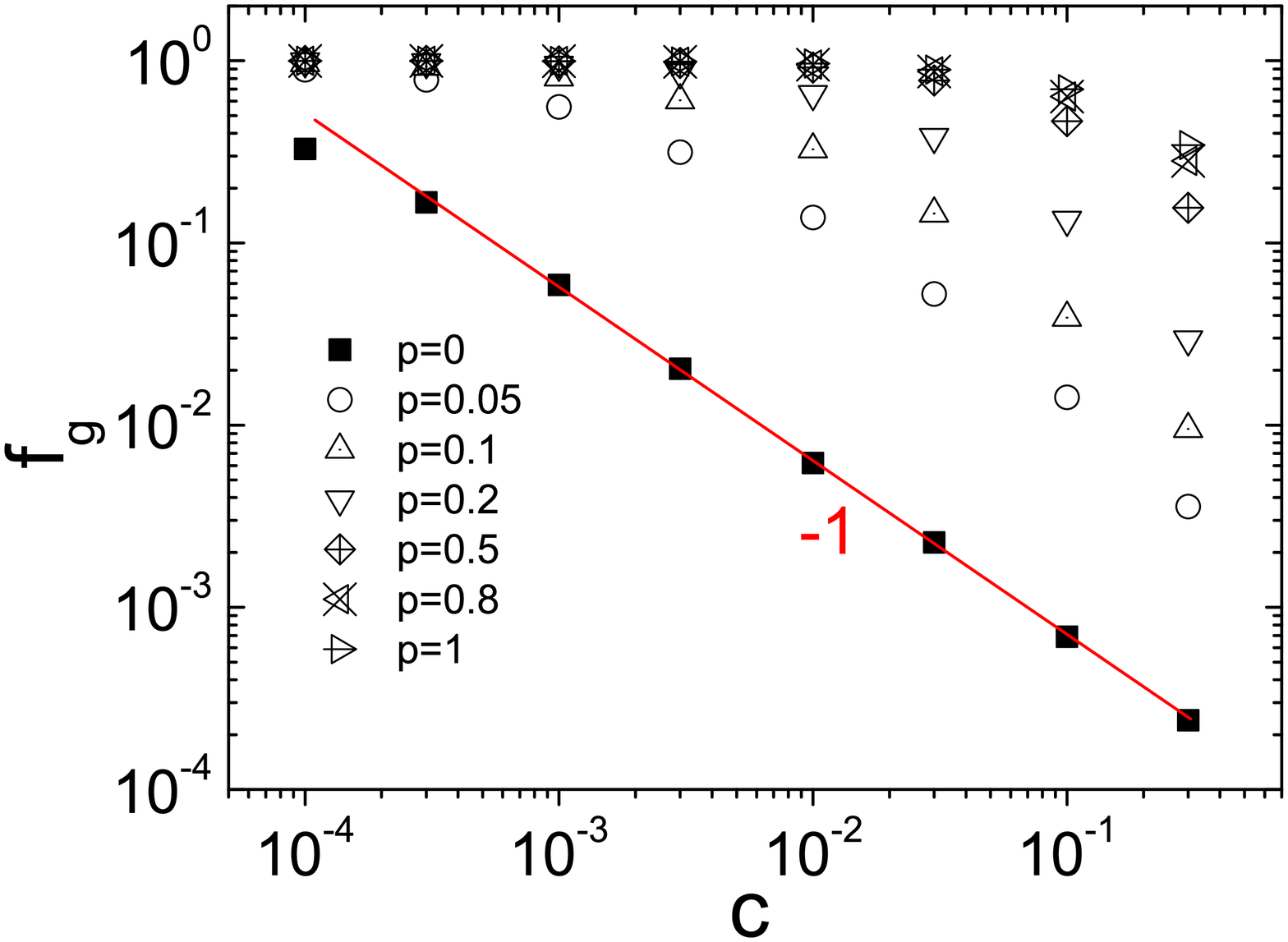}
    \label{SFfg1:a}}
    \subfigure[]{
    \includegraphics[width=7cm]{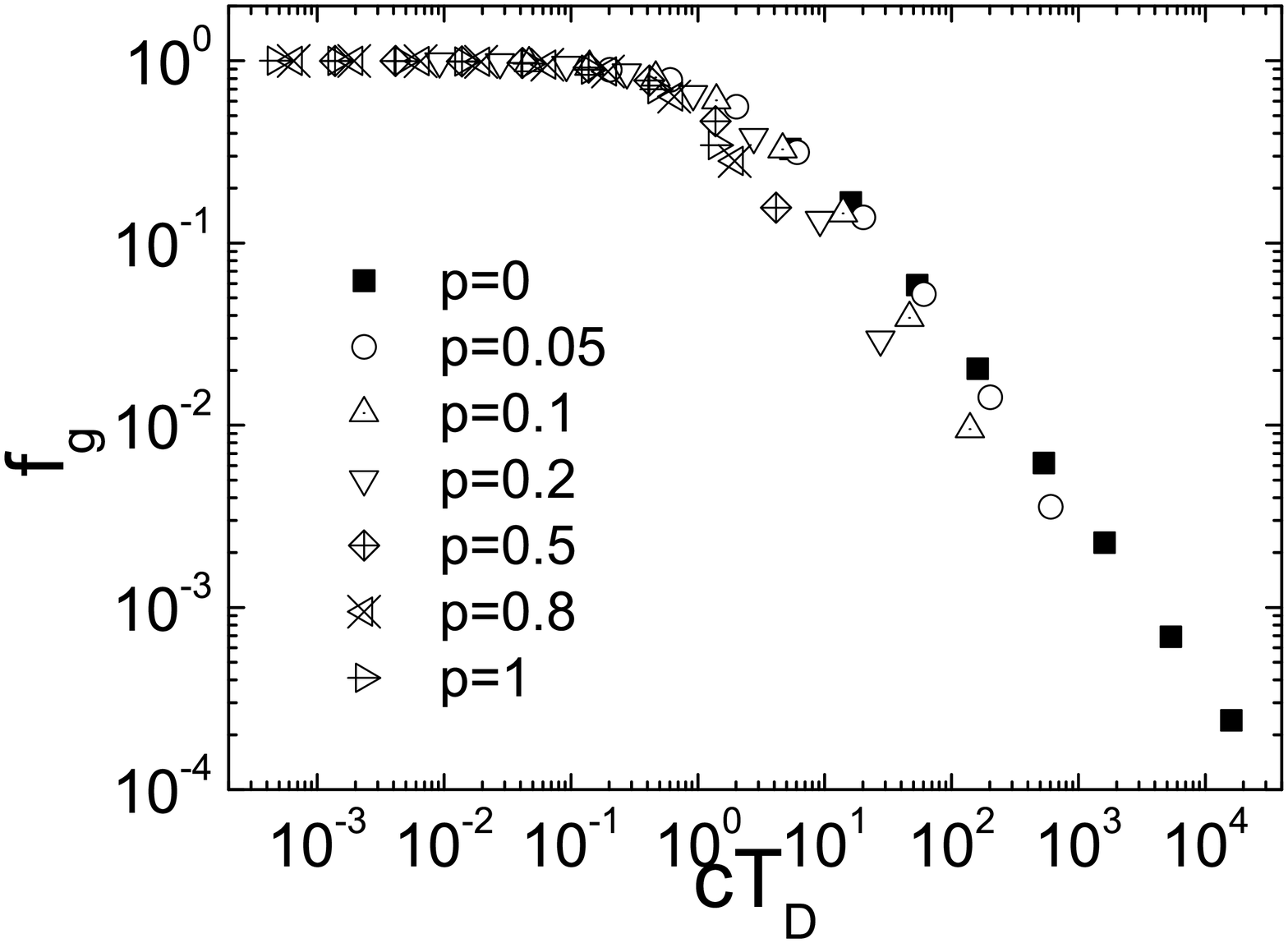}
    \label{SFfg1:b}}
    \end{center}
		\caption{(Color online) (a) Log-log plot of $f_g$ vs $c$ for various values of $p$ for SF networks with $\gamma=2.5$ and $N=20000$.  (b) Log-log plot of $f_g$ vs $c T_D$ for the same SF networks where $T_D$ is the MFPT for the case of no traps. Results are for different trap concentration $c$, $N=20000$ and for various values of $p$.}
\label{SFfg1}		
\end{figure}

We also study $f_g$ for scale free networks with degree distribution 
$p(k) \sim k^{-\gamma}$. We find in SF networks similar behavior, see Fig. \ref{SFfg1}, and there is a similar abrupt decay of $f_g$ above $c_{th}$. However, in this case the MFPT is not known analytically for every value of $p$. Instead, it is possible to use the upper bound value of the MFPT \cite{skarpa_pre}. It will not give the minimum value $p_m$, but a certainly secured value  with applicability to real networks. Our results on SF suggest that it is possible to secure almost all the messages by only applying a minimum, often very small, bias. What is needed for this is a minimal knowledge of the structure of the network or a strategy to increase the probability to get closer to the target at each step. In comparison with ER networks, we see that in SF networks a smaller bias is needed to have a significant improvement in $f_g$. This is mainly due to the MFPT, which is significantly smaller in a SF networks compared to ER networks when a bias towards the target is applied \cite{skarpa_pre}. This is since, due to hubs, the distances between sources and targets are smaller on the order of $\log(\log{N})$ \cite{ultrasmall} compared to ER for which the distances scale as $\log{N}$ \cite{Bollobas_RandGraph}.

\begin{figure}
    \begin{center}
\includegraphics[width=8cm]{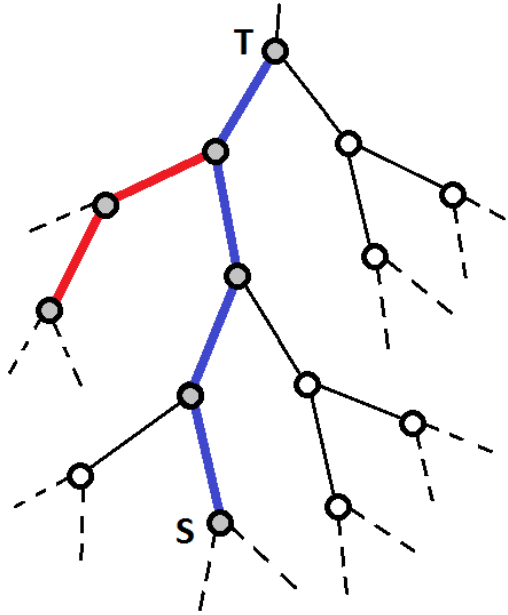}
\end{center}
\caption{(Color online) For a sufficiently large value of $p$, we assume that the mean number of distinct nodes $S$ can be approximated by $D$(blue)$+\frac{1}{2}(T_D-D)$(red)$=\frac{1}{2}(T_D+D)$. }
\label{sfig}
\end{figure}

\begin{figure}
    \begin{center}
\includegraphics[width=12cm]{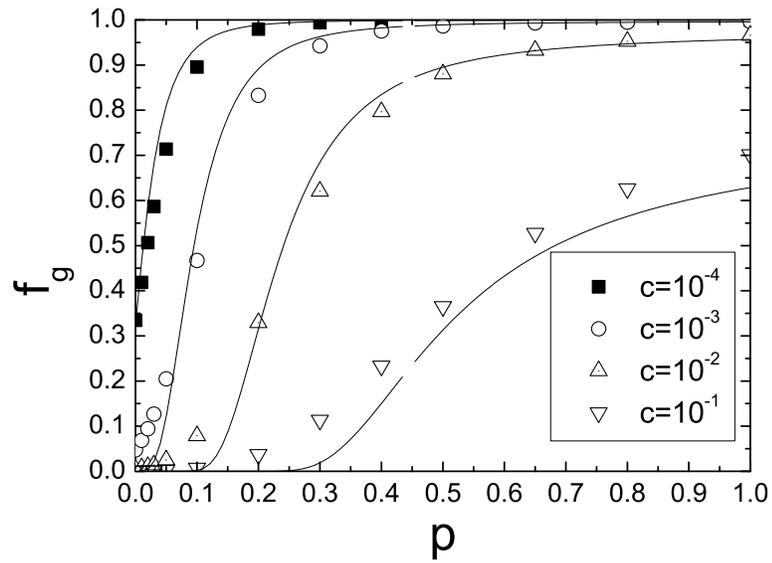}
\end{center}
\caption{Linear plot of $f_g$ vs $p$ for different $c$ values in ER networks with $\av{k}=10$ and $N=20000$. The symbols represent simulation results while the lines corresponds to the approximated analytic expression Eq. (\ref{eq:approx}).}
\label{fgvsp}

\end{figure}

According to Eq. (\ref{eq:fgexact}), the key quantity is the distribution of the number of distinct nodes $S$ visited before meeting the target. Even if it is difficult to find an analytic expression of the function $q_S$ for all cases, we can still try to find an approximation for a sufficiently large value of the bias parameter, $p$, since in this case the distribution of $S$ is not so broad. In this case, we can find an approximation of the mean value of $S$ by considering the typically tree-like structure of ER networks (see Fig. \ref{sfig}). When the particle deviates from the shortest path, we assume that it passes twice the same nodes and the mean number of distinct nodes visited can then be approximated by: 
\begin{equation} 
S\simeq D+\frac{1}{2}(T_D-D)=\frac{1}{2}(T_D+D),
\end{equation}
where $T_D$ and $D$ are the known MFPT and mean shortest distance in the network, respectively. 
Thus, for large values of $p$ (e.g. $p>0.5$), $f_g$ can be approximated by the following expression:
\begin{equation} 
f_g\simeq (1-c)^{\frac{1}{2}(T_D+D)}
\label{eq:approx}
\end{equation}
Since we know analytically both $D$ and $T_D$ for ER networks, it is possible to test the validity of this approximation. In Fig. \ref{fgvsp}, we compare simulation results (symbols) and the corresponding approximation, Eq (\ref{eq:approx}). Thus, we see that a very simple argument gives a surprisingly good analytic approximation, even for relatively small bias.

\begin{figure}[htbp]
    \begin{center}
    \includegraphics[width=12cm]{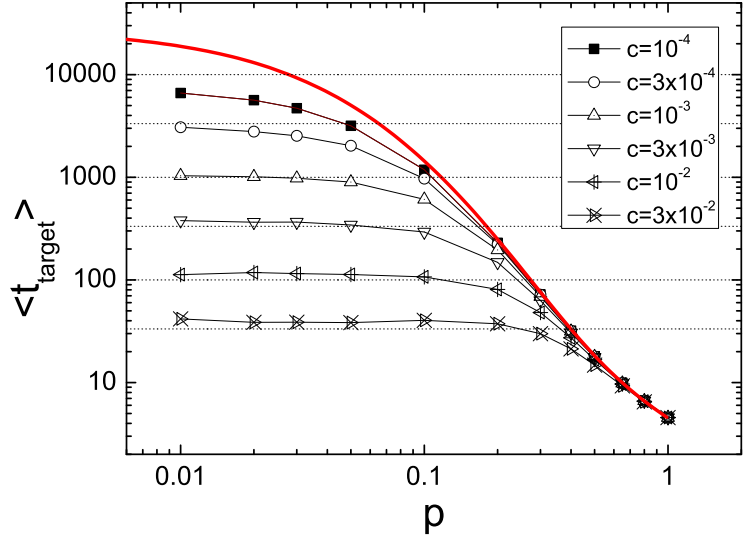}
    \end{center}
		\caption{(Color online) $\av{t_{target}}$ vs $p$ for ER networks with $\av{k}=10$, $N=20000$, for various values of $c$. The dotted lines correspond to the $1/c$ values and the red curve to the $p$ dependence of the MFPT without traps according to Eq. (\ref{eq:solbis}).}
\label{ER_tp}		
\end{figure}

It is also useful to investigate the dependence of the average time of the survived particles to reach the target, $\av{t_{target}}$, on $p$ in the presence of traps. In Fig. \ref{ER_tp} we analyze $\av{t_{target}}$ and find that the bias $p$ has a significant effect on the average time, decreasing it, only for values of $p$ above a certain threshold, which depends on the concentration of traps (higher values for higher concentrations). When the bias is low, the process is controlled by the presence of traps and $\av{t_{target}}\sim 1/c$, see dotted lines in Fig. 6. This can be understood by the fact that the target is a kind of trap, and when the bias is not significant, the targets become indistinguishable from the traps. In this case the distance between the traps is the important parameter controlling the diffusion time. On the other hand, when the bias becomes important the average time to diffuse to the target becomes less than $1/c$ and the process is controlled by the dependence of the MFPT on the bias parameter $p$ (red line in Fig. \ref{ER_tp}).  
 
\section{Single trap - SF networks}

\begin{figure}[htbp]
    \begin{center}
    \subfigure[]{
    \includegraphics[width=7cm]{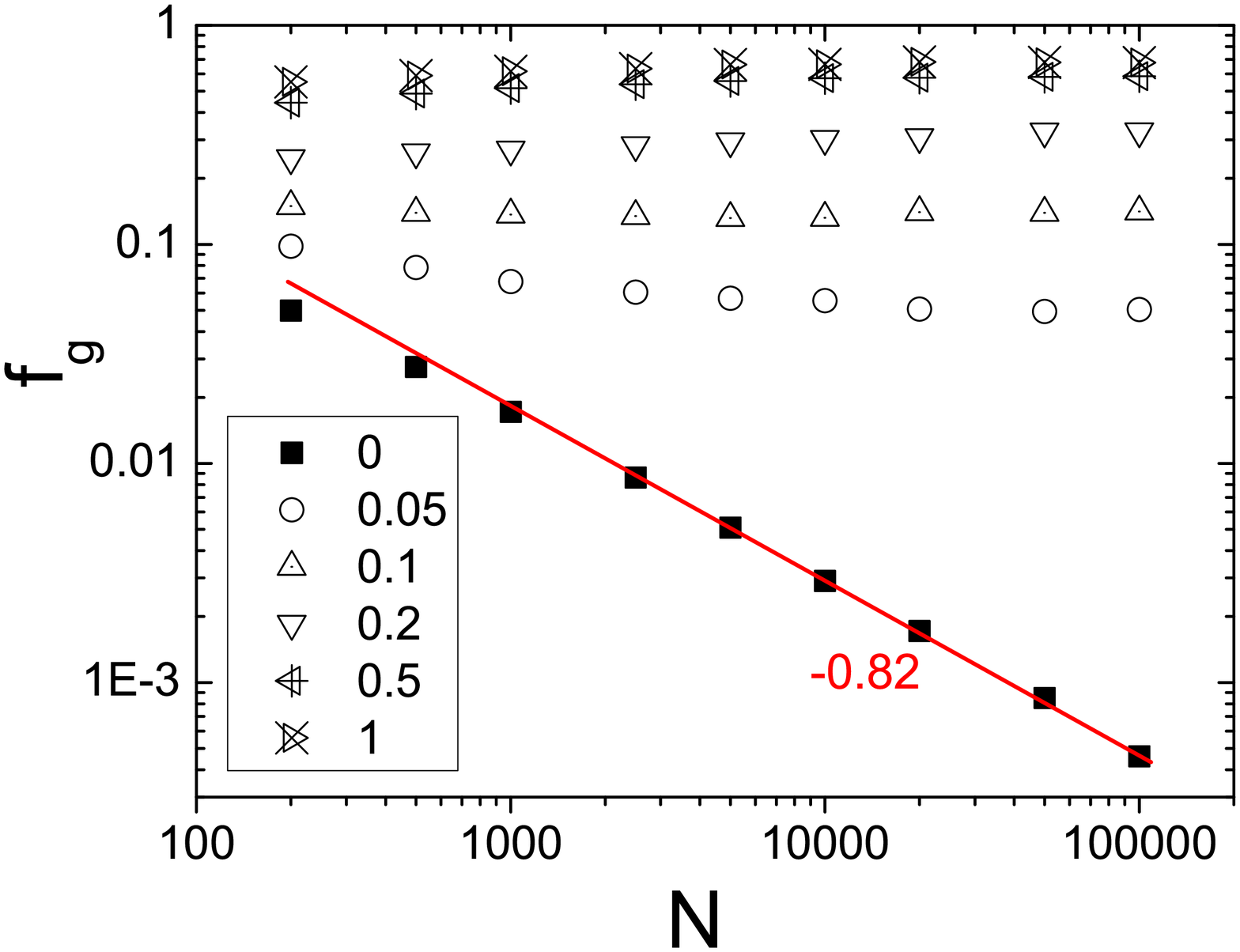}
    \label{SF_fgN:a}}
    \subfigure[]{
    \includegraphics[width=7cm]{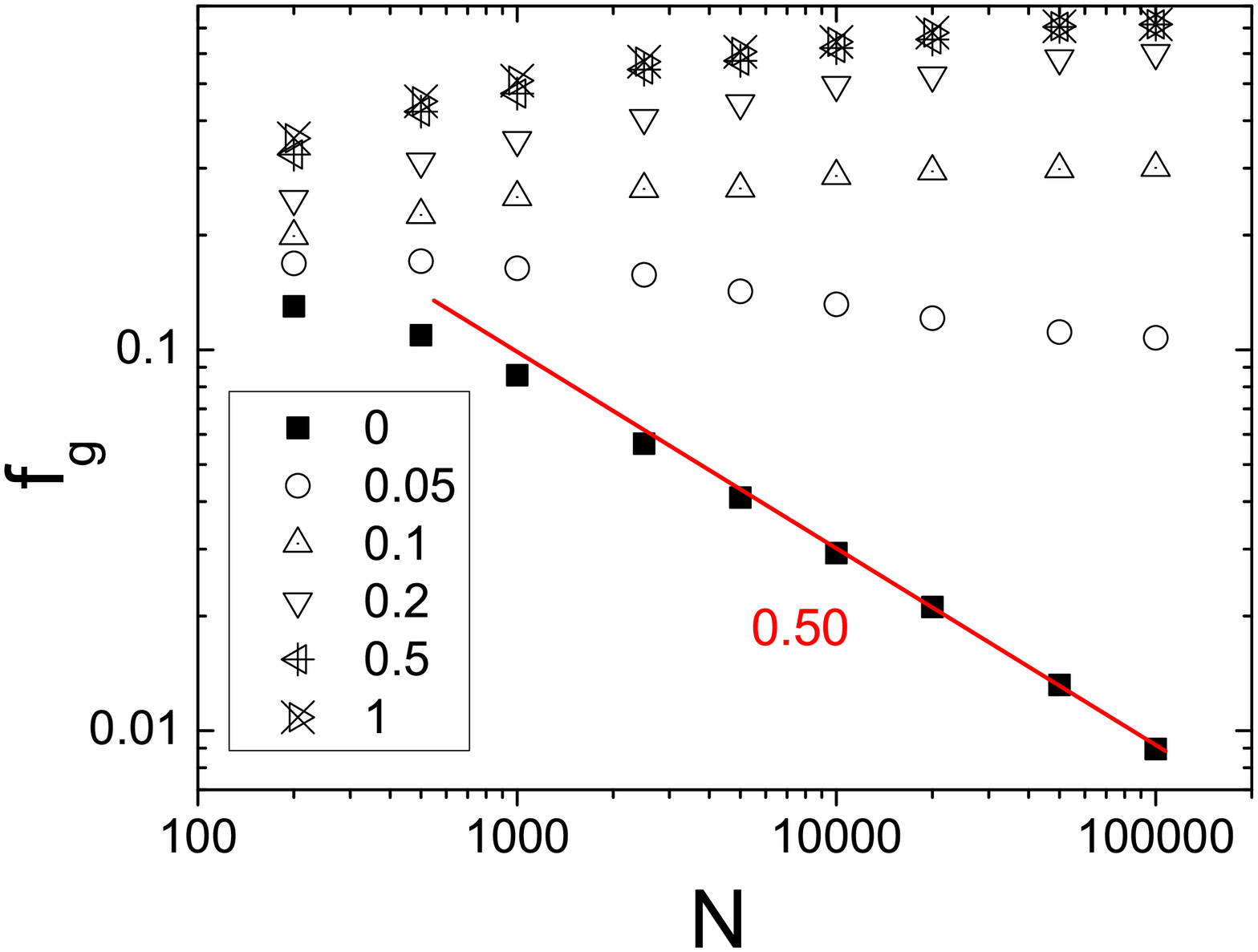}
    \label{SF_fgN:b}}
    \end{center}
		\caption{(Color online) Log-log plot of $f_g$ vs $N$ when the largest hub is a trap for SF networks with  a) $\gamma=2.2$ and b) $\gamma=3$. Note that the slopes for the unbiased case agree well with Eq. (\ref{eq:hub2}).}
\label{SF_fgN}		
\end{figure}

\begin{figure}
\begin{center}
\includegraphics[width=12cm]{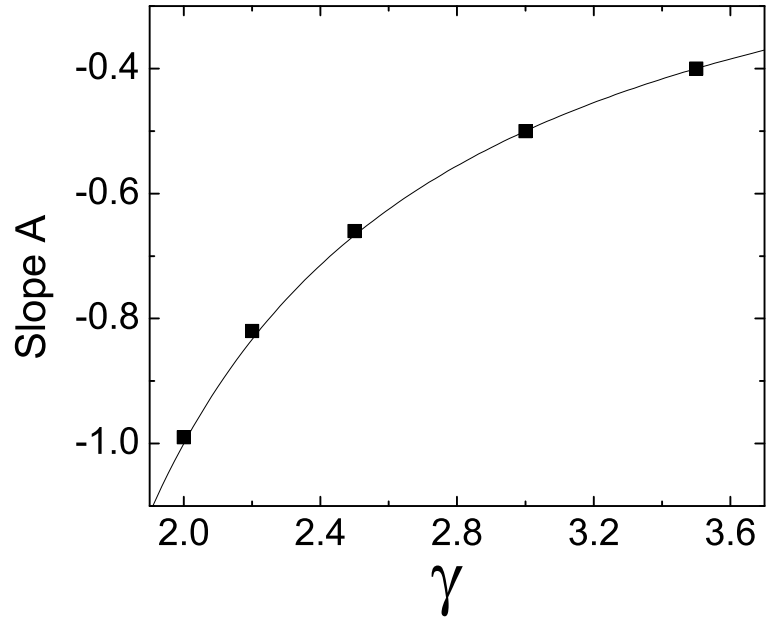}
\end{center}
\caption{Eq. (\ref{eq:hub2}) suggests that $\log{f_g} \sim A \log{N}$ where $A=-1/(\gamma-1)$. The line represents the theoretical $A=-1/(\gamma-1)$. The symbols are slopes from simulations of unbiased case ($p=0$) obtained from figures like \ref{SF_fgN:a} and \ref{SF_fgN:b}.}
\label{SF_slope}
\end{figure}

We are now interested in a scenario where a central hub of the network fails, as may be the case with real world networks since this kind of nodes is very prone to attacks. In this case, we assume that a single trap is placed on the highest degree node, $k_{max}$ of the network. When only one trap is present on the highest connected node of the network, we can use the following simple argument to evaluate $f_g$. We consider that all the $k_{max}$ nodes that are connected with the hub act like traps since they drive the particle to the trap. Thus, for the unbiased case, we can use $k_{max}/N$ instead of $c$ in equation (\ref{eq:fgrw}). Thus, we assume,
\begin{equation} 
f_g\sim \frac{1}{k_{max}}.
\label{eq:hub1}
\end{equation}
The scaling of $k_{max}$ with the size of the network is known for all $\gamma$ values to be $k_{max} \sim N^{\frac{1}{\gamma-1}}$ \cite{Cohen_resilience}. Thus, for the random walk case, without bias, we have:
\begin{equation} 
f_g \sim N^{-\frac{1}{\gamma-1}}
\label{eq:hub2}
\end{equation} 
In Figs. \ref{SF_fgN:a} and \ref{SF_fgN:b} we see the results for the scaling of $f_g$ with system size $N$ for various $p$ values in SF networks with $\gamma =2.2$ and 3, respectively, when a single trap is present on the most connected node. For the unbiased walk the dependance of the slope on $\gamma$ is in very good agreement with Eq. (\ref{eq:hub2}) (see also Fig. \ref{SF_slope}). This relation  shows the increasing role of the hub when $\gamma$ is decreased. For small $\gamma$, most particles traversing the network need to pass through the hub and are subsequently trapped while for higher $\gamma$ values, the particle is more likely to find the target without passing from through the hub. When considering a biased random walk, the general picture from figures \ref{SF_fgN:a} and \ref{SF_fgN:b} is that when the size of the network becomes sufficiently large, $f_g$ remains almost unaffected by system size for every value of $p$. This means that in a SF network with failure of the most connected node, even a small bias can vastly improve the transmission process in comparison with the unbiased case, and the effect is more pronounced the larger the network. In SF network with large $\gamma$ values, the effect is less pronounced, however, a small bias still offers a significant improvement in the fraction of particles that successfully arrive on target.

\section{Conclusions}
We propose a model to study the efficiency of biased random walks in ER and SF networks to reach their targets in the presence of traps. We find that in the presence of a concentration $c$ of randomly distributed traps, for the unbiased random walk, the dependence of the fraction of particles that arrive on target $f_g$ on $c$ is $f_g\sim c^{-1}$ in both ER and SF networks. For biased walks, there exists a threshold value of the concentration $c_{th}$ which depends on the bias $p$. For $c<c_{th}$ almost all the particles arrive on target without being trapped i.e $f_g\approx 1$. For $c>c_{th}$,  $f_g$ decreases significantly approximately as $1/c$. This threshold value, $c_{th}$, corresponds to the inverse of the MFPT of the biased diffusion process without traps, and thus it is possible for any concentration of traps to find the needed minimum bias parameter $p$ in order to have unaffected transmission. Also, by a simple approximation of the mean numbers of distinct nodes visited, we obtain an analytic expression that adequately describes the function $f_g$ for a broad range of values of $p$. We have also investigated the efficiency of the process in SF networks after failure of the most connected node which behave as a trap. In the case of the unbiased random walk, we find that $f_g \sim 1/ k_{max}$ and thus $f_g \sim N^{-1/(\gamma-1)}$. In the case of biased walks, even a small bias can vastly improve the efficiency of the transmission process in comparison with the unbiased case. The improvement is more pronounced the larger is the size of the network, since the effective concentration of traps, $k_{max}/N$, decreases with $N$.

\begin{acknowledgments}
Aknowledgements: This research has been co-financed by the European Union (European Social Fund - ESF) and Greek national funds through the Operational Program ``Education and Lifelong Learning" of the National Strategic Reference Framework (NSRF) - Research Funding Program: Heracleitus II (to LS). SH wishes to thank the LINC EU project and the EU-FET project MULTIPLEX 317532, DTRA, ONR, the DFG and the Israel Science Foundation for support.
\end{acknowledgments}

\bibliography{netbias}

\begin{thebibliography}{16}
\expandafter\ifx\csname natexlab\endcsname\relax\def\natexlab#1{#1}\fi
\expandafter\ifx\csname bibnamefont\endcsname\relax
  \def\bibnamefont#1{#1}\fi
\expandafter\ifx\csname bibfnamefont\endcsname\relax
  \def\bibfnamefont#1{#1}\fi
\expandafter\ifx\csname citenamefont\endcsname\relax
  \def\citenamefont#1{#1}\fi
\expandafter\ifx\csname url\endcsname\relax
  \def\url#1{\texttt{#1}}\fi
\expandafter\ifx\csname urlprefix\endcsname\relax\def\urlprefix{URL }\fi
\providecommand{\bibinfo}[2]{#2}
\providecommand{\eprint}[2][]{\url{#2}}

\bibitem[{\citenamefont{Weiss}(1994)}]{weiss_random}
\bibinfo{author}{\bibfnamefont{G.~H.} \bibnamefont{Weiss}},
  \emph{\bibinfo{title}{Aspects and applications of the random walk}}
  (\bibinfo{publisher}{North-Holland}, \bibinfo{address}{Amsterdam},
  \bibinfo{year}{1994}).

\bibitem[{\citenamefont{Hollander and Weiss}(2001)}]{hollander_problems}
\bibinfo{author}{\bibfnamefont{F.}~\bibnamefont{Hollander}} \bibnamefont{and}
  \bibinfo{author}{\bibfnamefont{G.~H.} \bibnamefont{Weiss}},
  \emph{\bibinfo{title}{Contemporary problems in statistical physics}}
  (\bibinfo{publisher}{MIT Press}, \bibinfo{address}{Cambridge},
  \bibinfo{year}{2001}).

\bibitem[{\citenamefont{Bunde et~al.}(1997)\citenamefont{Bunde, Havlin,
  Klafter, Graff, and Shehter}}]{bunde_prl78}
\bibinfo{author}{\bibfnamefont{A.}~\bibnamefont{Bunde}},
  \bibinfo{author}{\bibfnamefont{S.}~\bibnamefont{Havlin}},
  \bibinfo{author}{\bibfnamefont{J.}~\bibnamefont{Klafter}},
  \bibinfo{author}{\bibfnamefont{G.}~\bibnamefont{Graff}}, \bibnamefont{and}
  \bibinfo{author}{\bibfnamefont{A.}~\bibnamefont{Shehter}},
  \bibinfo{journal}{Phys. Rev. Lett.} \textbf{\bibinfo{volume}{78}},
  \bibinfo{pages}{3338} (\bibinfo{year}{1997}).

\bibitem[{\citenamefont{Havlin et~al.}(1990)\citenamefont{Havlin, Larralde,
  Kopelman, and Weiss}}]{havlin_physa169}
\bibinfo{author}{\bibfnamefont{S.}~\bibnamefont{Havlin}},
  \bibinfo{author}{\bibfnamefont{H.}~\bibnamefont{Larralde}},
  \bibinfo{author}{\bibfnamefont{R.}~\bibnamefont{Kopelman}}, \bibnamefont{and}
  \bibinfo{author}{\bibfnamefont{G.~H.} \bibnamefont{Weiss}},
  \bibinfo{journal}{Physica A} \textbf{\bibinfo{volume}{169}},
  \bibinfo{pages}{337} (\bibinfo{year}{1990}).

\bibitem[{\citenamefont{Donsker and Varadhan}(1979)}]{donsker_com32}
\bibinfo{author}{\bibfnamefont{N.~D.} \bibnamefont{Donsker}} \bibnamefont{and}
  \bibinfo{author}{\bibfnamefont{S.~R.~S.} \bibnamefont{Varadhan}},
  \bibinfo{journal}{Commun. Pure Appl. Math.} \textbf{\bibinfo{volume}{32}},
  \bibinfo{pages}{721} (\bibinfo{year}{1979}).

\bibitem[{\citenamefont{Jasch and Blumen}(2001)}]{jasch_pre64}
\bibinfo{author}{\bibfnamefont{F.}~\bibnamefont{Jasch}} \bibnamefont{and}
  \bibinfo{author}{\bibfnamefont{A.}~\bibnamefont{Blumen}},
  \bibinfo{journal}{Phys. Rev. E} \textbf{\bibinfo{volume}{64}},
  \bibinfo{pages}{066104} (\bibinfo{year}{2001}).

\bibitem[{\citenamefont{Kittas et~al.}(2008)\citenamefont{Kittas, Carmi,
  Havlin, and Argyrakis}}]{kittas_epl84}
\bibinfo{author}{\bibfnamefont{A.}~\bibnamefont{Kittas}},
  \bibinfo{author}{\bibfnamefont{S.}~\bibnamefont{Carmi}},
  \bibinfo{author}{\bibfnamefont{S.}~\bibnamefont{Havlin}}, \bibnamefont{and}
  \bibinfo{author}{\bibfnamefont{P.}~\bibnamefont{Argyrakis}},
  \bibinfo{journal}{Europhys. Lett.} \textbf{\bibinfo{volume}{84}},
  \bibinfo{pages}{40008} (\bibinfo{year}{2008}).

\bibitem[{\citenamefont{Kittas and Argyrakis}(2009)}]{kittas_pre80}
\bibinfo{author}{\bibfnamefont{A.}~\bibnamefont{Kittas}} \bibnamefont{and}
  \bibinfo{author}{\bibfnamefont{P.}~\bibnamefont{Argyrakis}},
  \bibinfo{journal}{Phys. Rev. E} \textbf{\bibinfo{volume}{80}},
  \bibinfo{pages}{046111} (\bibinfo{year}{2009}).

\bibitem[{\citenamefont{Gallos}(2004)}]{gallos_pre70}
\bibinfo{author}{\bibfnamefont{L.~K.} \bibnamefont{Gallos}},
  \bibinfo{journal}{Phys. Rev. E} \textbf{\bibinfo{volume}{70}},
  \bibinfo{pages}{046116} (\bibinfo{year}{2004}).

\bibitem[{\citenamefont{Avin and Brito}(2004)}]{avin_query}
\bibinfo{author}{\bibfnamefont{C.}~\bibnamefont{Avin}} \bibnamefont{and}
  \bibinfo{author}{\bibfnamefont{C.}~\bibnamefont{Brito}}, in
  \emph{\bibinfo{booktitle}{Proc. of the third international symposium on
  Information processing in sensor networks}} (\bibinfo{year}{2004}), pp.
  \bibinfo{pages}{277--286}.

\bibitem[{\citenamefont{Bar-Yossef et~al.}(2006)\citenamefont{Bar-Yossef,
  Friedman, and Kliot}}]{yossef_mobi}
\bibinfo{author}{\bibfnamefont{Z.}~\bibnamefont{Bar-Yossef}},
  \bibinfo{author}{\bibfnamefont{R.}~\bibnamefont{Friedman}}, \bibnamefont{and}
  \bibinfo{author}{\bibfnamefont{G.}~\bibnamefont{Kliot}}, in
  \emph{\bibinfo{booktitle}{MobiHoc '06: Proceedings of the seventh ACM
  international symposium on Mobile ad hoc networking and computing}}
  (\bibinfo{publisher}{ACM Press}, \bibinfo{address}{New-York, NY, USA},
  \bibinfo{year}{2006}), pp. \bibinfo{pages}{238--249}.

\bibitem[{\citenamefont{Gkantsidis et~al.}(2004)\citenamefont{Gkantsidis,
  Mihail, and Saberi}}]{gkantsidis_peer}
\bibinfo{author}{\bibfnamefont{C.}~\bibnamefont{Gkantsidis}},
  \bibinfo{author}{\bibfnamefont{M.}~\bibnamefont{Mihail}}, \bibnamefont{and}
  \bibinfo{author}{\bibfnamefont{A.}~\bibnamefont{Saberi}}, in
  \emph{\bibinfo{booktitle}{Proc. 23 Annual Joint Conference of the IEEE
  Computer and Communications Societies (INFO-COM)}}
  (\bibinfo{publisher}{Elsevier Science Publishers B. V.},
  \bibinfo{year}{2004}).

\bibitem[{\citenamefont{Skarpalezos et~al.}(2013)\citenamefont{Skarpalezos,
  Kittas, Argyrakis, Cohen, and Havlin}}]{skarpa_pre}
\bibinfo{author}{\bibfnamefont{L.}~\bibnamefont{Skarpalezos}},
  \bibinfo{author}{\bibfnamefont{A.}~\bibnamefont{Kittas}},
  \bibinfo{author}{\bibfnamefont{P.}~\bibnamefont{Argyrakis}},
  \bibinfo{author}{\bibfnamefont{R.}~\bibnamefont{Cohen}}, \bibnamefont{and}
  \bibinfo{author}{\bibfnamefont{S.}~\bibnamefont{Havlin}},
  \bibinfo{journal}{Phys. Rev. E} \textbf{\bibinfo{volume}{88}},
  \bibinfo{pages}{012817} (\bibinfo{year}{2013}).

\bibitem[{\citenamefont{Cohen and Havlin}(2003)}]{ultrasmall}
\bibinfo{author}{\bibfnamefont{R.}~\bibnamefont{Cohen}} \bibnamefont{and}
  \bibinfo{author}{\bibfnamefont{S.}~\bibnamefont{Havlin}},
  \bibinfo{journal}{Phys. Rev. Lett.} \textbf{\bibinfo{volume}{90}},
  \bibinfo{pages}{058701} (\bibinfo{year}{2003}).

\bibitem[{\citenamefont{Bollob\'{a}s}(2001)}]{Bollobas_RandGraph}
\bibinfo{author}{\bibfnamefont{B.}~\bibnamefont{Bollob\'{a}s}},
  \emph{\bibinfo{title}{Random Graphs}} (\bibinfo{publisher}{Cambridge
  University Press}, \bibinfo{address}{New York}, \bibinfo{year}{2001}).

\bibitem[{\citenamefont{Cohen et~al.}(2000)\citenamefont{Cohen, Erez, ben
  Avraham, and Havlin}}]{Cohen_resilience}
\bibinfo{author}{\bibfnamefont{R.}~\bibnamefont{Cohen}},
  \bibinfo{author}{\bibfnamefont{K.}~\bibnamefont{Erez}},
  \bibinfo{author}{\bibfnamefont{D.}~\bibnamefont{ben Avraham}},
  \bibnamefont{and} \bibinfo{author}{\bibfnamefont{S.}~\bibnamefont{Havlin}},
  \bibinfo{journal}{Phys. Rev. Lett.} \textbf{\bibinfo{volume}{85}},
  \bibinfo{pages}{4626} (\bibinfo{year}{2000}).

\end{thebibliography}



\end{document}